\documentclass[11pt]{article}
\usepackage{graphics}
\def \lp{\scriptscriptstyle +}

\def\sss{\rm_S}
\def\phrac#1/#2{\leavevmode\kern.1em\raise.5ex\hbox{\the\scriptfont0 #1}
\kern-.1em/\kern-.15em\lower.25ex\hbox{\the\scriptfont0 #2}}
\begin{document}
\parskip 3pt plus 1pt
\baselineskip=12 pt plus 2pt minus 1pt
\null
\def\moriond{\vtop{\hbox{XXIII{\small{rd}} Rencontre de Moriond}
                   \hbox{Current Issues in Hadron Physics}
                   \hbox{Les Arcs, France (13--19 March 1988) 451--454}}}
\rightline{\moriond}
\vskip 60pt
\centerline {\bf CHARM \thinspace HADROPRODUCTION \thinspace AT \thinspace
FERMILAB \thinspace E769}
\vskip 12pt
\centerline {The TPL Collaboration}
\vskip 8pt \noindent
D.J. Summers, G.A. Alves, J.C. Anjos, J.A. Appel, M. Aryal, S.B. Bracker,
L.M.~Cremaldi, 
R.L.\,Dixon, D.\,Errede,\,H.C.\,Fenker,\,C.\,Gay, 
D.R.\,Green,
R.\,Jedicke,\,D.\,Kaplan, P.E.\,Karchin, \linebreak
I.\,Leedom, L.H.\,Lueking, 
G.J.\,Luste,
A.B. de Oliveira, P.M. Mantsch, S.~Marques, J.~Metheny, \linebreak
R. Milburn,
H. da Motta, A. Napier, T. Nash, S. Reucroft, A.F.S.~Santoro, M.~Sheaff, 
C.~Stoughton, M.H.G. Souza, W.J. Spalding, M.E. Streetman, X. Wu
\vskip 8pt
\centerline{CBPF--Rio de Janeiro, Fermilab, Northeastern, Toronto,
Tufts, Wisconsin, Yale}
\centerline{Work
supported by CNPq(Brazil), DOE(U.S.), NSERC(Canada) and NSF(U.S.)}
\vskip 12pt
\centerline {Presented by}
\vskip 4pt
\centerline {D. J. Summers}
\centerline {Fermilab}
\centerline {Batavia, Illinois 60510}
\centerline {United States}
\vspace*{7mm}
\leftline{\hspace{56mm}
\resizebox{40mm}{!}{\includegraphics{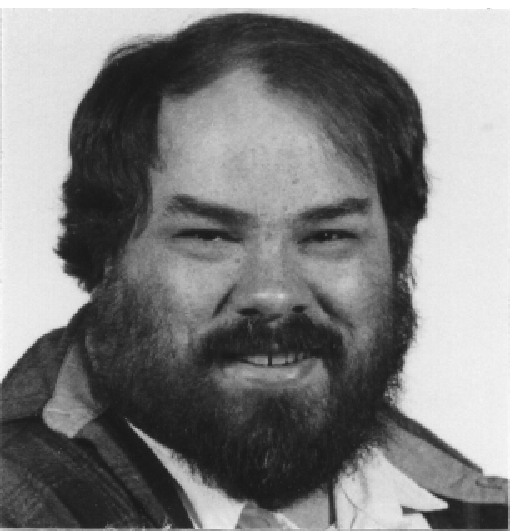}}}
\vspace*{7mm}

{\bf ABSTRACT}

E769 has just recorded on tape the interactions of 500 million
pions, kaons, and protons.
A \v Cerenkov counter and
a TRD were used to
tag beam particle types in both positive and
negative 250 GeV/c hadron beams.  Thin foil Be, Al, Cu, and W targets
were used with a spectrometer including silicon microstrips
to look for charm decay vertices.
Preliminary results show
$D^0 \rightarrow K^-\pi^+$ and $D^+ \rightarrow K^-\pi^+\pi^+$
mass peaks. As the event reconstruction progresses on numerous
parallel microprocessors,
we intend to explore the $p_{_T}$, $x_F$, A, and
flavor dependence of the production of charmed mesons and baryons.

\eject
\leftline{\bf INTRODUCTION}

   Experiment 769 was designed to explore how pions, kaons, and protons produce
charmed mesons and baryons.  To achieve this goal we modified and
upgraded the Tagged Photon Laboratory (TPL) at
Fermilab.  Previously, this apparatus had been used by E691 [1] to \linebreak
photoproduce and fully reconstruct over 10,000 charmed particles.
Silicon microstrip planes \linebreak
are employed to separate the primary vertex in an
event from secondary charm decay vertices.  We used a fast data acquisition
system to record a large quantity of data with a fairly open \linebreak
global E$_{_T}$\,trigger.~We rely on offline
vertexing, mass reconstruction, and particle identification \linebreak
to \, find \, charm. Production event filtering and reconstruction will soon
be ready to start on \linebreak
an ACP [2] microprocessor farm.

\vskip 10pt
\leftline{\bf BEAM PARTICLE IDENTIFICATION AND TRACKING}

   Data was taken with both positive and negative 250 GeV/c hadron beams
at maximum rates of 2 to 4 MHz.
The positive beam was a mixture of
approximately 59\% $\pi^+$, 35\% $p$, and \linebreak
6\% $K^+$
and the negative beam 91\% $\pi^-$, 2\% $\overline{p}$,
and 7\% $K^-$.

\hangindent=-83truemm
\hangafter=2
   To explore kaon production of the $D_{\sss}^{\lp}$ and other
charmed particles, it was necessary to identify and trigger on
kaons to increase the number of recorded kaon interactions. This was \linebreak
done
with a Differential Isochronous Self--Focusing \v Cerenkov Counter (DISC)
developed at CERN and
Fermilab.
The DISC helium gas pressure was adjusted so that
\v Cerenkov light from a desired particle type would be focused onto a
narrow annular slit \linebreak
with adjustable width.  Eight phototubes sensed this light.
The counter was capable \linebreak
of resolving the 0.069 milliradian
difference in the opening half angle of pion and kaon light cones.  Figure 1
shows the number of times 7 or 8 phototubes were above threshold as a
function of pressure during a calibration run.
For data taking, at least one phototube in each of 4 quadrants
was required and the annular slit was made somewhat wider.
Pion contamination of kaon triggers is less than 10\%.

\hangindent=-83truemm
\hangafter=-100
   To differentiate between $\pi^+$ and $p$ \linebreak
beam particles, we built a
Transition Ra\-diation Detector (TRD). It consisted of 24 stacks of 200
12.7$\mu$ sheets of polypropylene radiators, $(CH_2)_n$.
Each stack was spaced over 50mm and was followed
by two xenon gas proportional wire chamber (PWC) planes to detect x--rays.
Figure~2 shows the distribution of the number
of PWC planes per event
which
were above threshold during a run for non--kaon triggers.
Note the proton peak at 5 and the pion peak
at 19. Pion and proton events with another beam particle arriving 
within \linebreak
150~ns were vetoed to improve particle~iden\-tification.

\vspace*{-147mm}
\leftline{\hspace{70mm}
\resizebox{88mm}{!}{\includegraphics{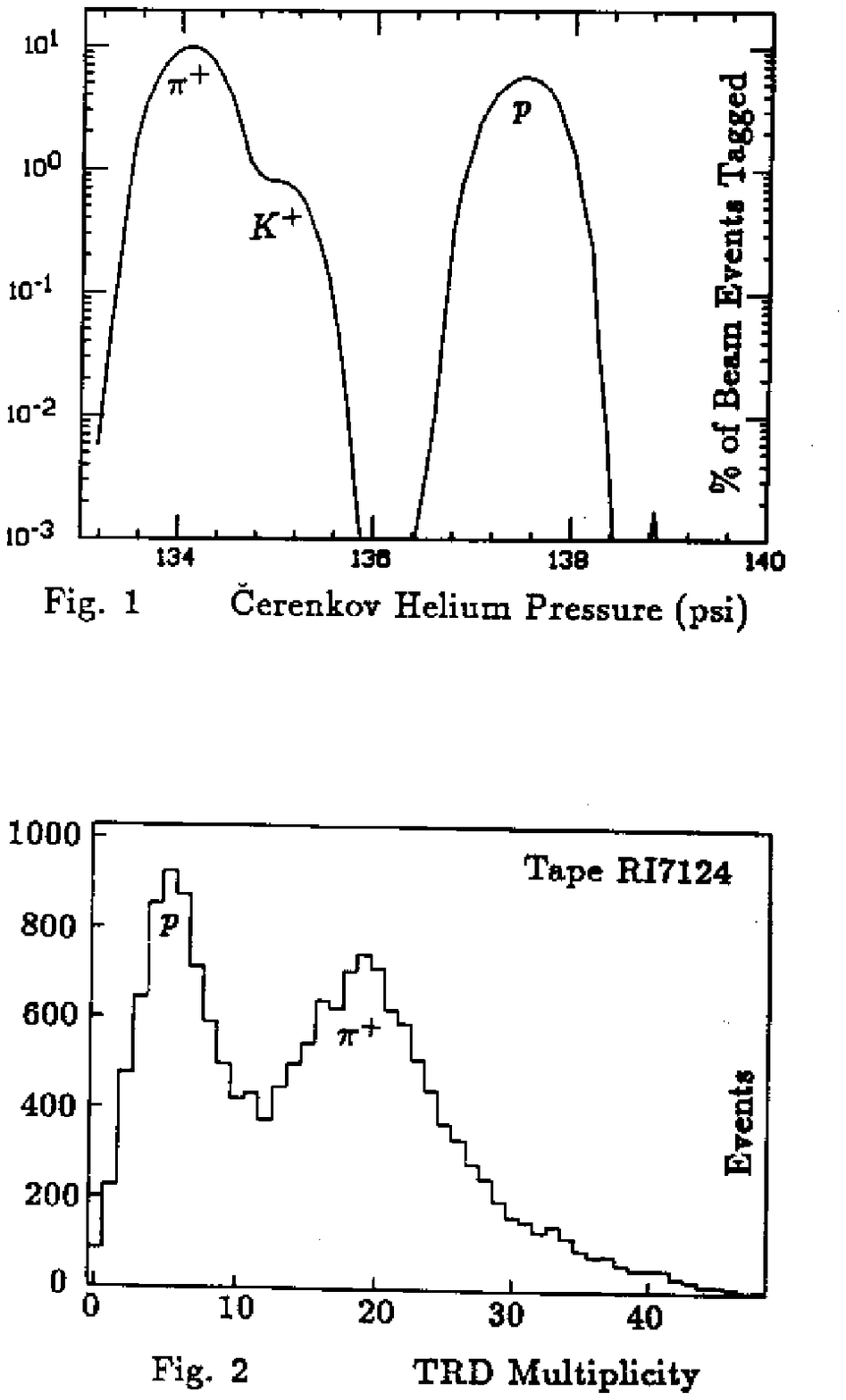}}}
\eject

   A system of eight 1mm pitch PWC planes far upstream of the target
and two 25$\mu$ pitch
silicon microstrip planes just upstream of the target were used to find
the beam track.  This yields a line pointing to the
primary vertex.
\vskip 8pt
\leftline{\bf FOIL TARGETS, SILICON MICROSTRIPS, AND SPECTROMETER}
Twenty-six thin foil targets
spaced at 1.7 mm intervals
($4\times100\mu$~W, $3\times250\mu$~Cu, $5\times250\mu$~Al, and
$14\times250\mu$~Be) were
used to {\it quantize} the {\it z}
position of primary interactions and to explore A~dependence. Both primary
interactions and downstream charm decays were
examined with 2~planes of 25$\mu$ pitch silicon
microstrip detectors and 9 planes of 50$\mu$ pitch~[3].
The downstream spectrometer [4] has 35 planes
of drift chambers, two PWC planes (2mm pitch), two momentum analyzing magnets,
two \v Cerenkov
counters, electromagnetic plus hadronic calorimeters,
and a plane of scintillators for muon detection.
\vskip 8pt
\leftline{\bf TRIGGER AND FAST DATA ACQUISITION}
\par
A loose global transverse energy (E$_{_T}$) trigger
enhanced the number of charm events. A second high E$_{_T}$ trigger further
enhanced total recorded charm, but at the expense of low $p_{_T}$ charm events.
All but DISC--tagged kaon interaction triggers were prescaled to enrich
the kaon fraction on tape.
\par
The data acquisition system [5], which reads out up to 400 4 \negthinspace{KB}
events/sec, is based on seven {\it smart} CAMAC crate
controllers (SCCs) connected in parallel to seven VMEbus double memory buffers
(RBUFs). A farm of
Motorola 68020 microprocessor boards, also in VMEbus, processed events.
The event processor board and some of the associated software \linebreak
were
designed by Fermilab's Advanced Computing
Project (ACP)~[6]. \, Because of the dou\-ble buffering, an SCC could
be reading one event from a CAMAC crate into an RBUF at 0.6~\negthinspace
$\mu$s/word; while an
event processor was extracting the previous event from the same 
RBUF. \linebreak
Data was written directly to a 6250 bpi tape drive
at 600 KB/s using a Ciprico VMEbus tape \linebreak
controller.  Two MB of memory on each
of 16 ACP boards was used to buffer events taken during a 22s spill.  This
allowed continuous tape writing during a 56s spill cycle.
\par
   The data run lasted from June 1987 until February 1988. Five hundred million
events were recorded on 10,000 tapes.  These included 180 million $\pi^-$,
25 million $K^-$,
100~million~$\pi^+$, 70 million $K^+$, and 70 million $p$ induced events.
\vskip 8pt
\leftline{\bf OFFLINE RECONSTRUCTION, CHARM PEAKS, AND PHYSICS GOALS}
\par
   To analyze 500 million events, we are using a farm of 55 ACP processors
based on \linebreak 
16\,MHz Motorola 68020s and 
programmed 
in FORTRAN\,[2].~The computing power is equal to \linebreak
40~VAX~11/780s.  If all events were fully reconstructed, about three years of
{\it farm} time would be required.  To reduce this to one year or less,
we are devising a fast filter which demands the possibility of a
secondary vertex in an event.
\par
    Preliminary results
shown in Figure~3 display
$D^+ \rightarrow K^-\pi^+\pi^+$,
$D^0 \rightarrow K^-\pi^+$, and $D^{*+} \rightarrow D^0\pi^+
(D^0 \rightarrow K^-\pi^+$)
mass peaks.
Charge conjugates are implicitly included.
The $D^+$ result comes from 8 million events and the $D^0$ peaks come
from $1{1\over{2}}$ million events.
An incomplete reconstruction program employing 9 out of 13 silicon
microstrip planes was used.
The cut $\sigma(\Delta{z})$ measures the significance of the separation
between primary and secondary vertices.  For the events in Figure 3c,
the measured
$D^{*+} - D^0$
mass difference was within 3~MeV/c$^2$ of the accepted
value, 145.5 MeV/c$^2$.
\par
As analysis
progresses, we intend to explore a number of physics topics including:
\begin{list}{$\bullet$}{
\setlength{\leftmargin}{0.7cm} 
\setlength{\itemsep}{0pt}
\setlength{\topsep}{1pt}
}
\item{The cross section for charm production by pions, kaons, and protons;
       including $d{\sigma}/dp_{_T}$.}
\item{Measuring $x_F$ distributions and parameterizing the gluon structure
       functions of the projectiles in the context of the gluon fusion model.}
\item{Charm production A dependence. Given $\sigma \propto
       A^\alpha$ and A = (9, 27, 63.5, 183.8); find $\alpha$.}
\item{Leading effects (e.g. more forward
       $\pi^+(u\overline{d}) \rightarrow D^+(c\overline{d})$ than
       $\pi^+(u\overline{d}) \rightarrow D^-(\overline{c}d)$).}
\item{Explore $D_{\sss}^{\lp}$ production by kaons and $\Lambda_{\rm{c}}^+$
       production by protons.}
\item{Exploit target foils and 25$\mu$ silicon microstrips to measure
       $D_{\sss}^{\lp}$ and $\Lambda_{\rm{c}}^+$ lifetimes.}
\end{list}
\break
\leftline{\hspace{-2mm}
\resizebox{158mm}{!}{\includegraphics{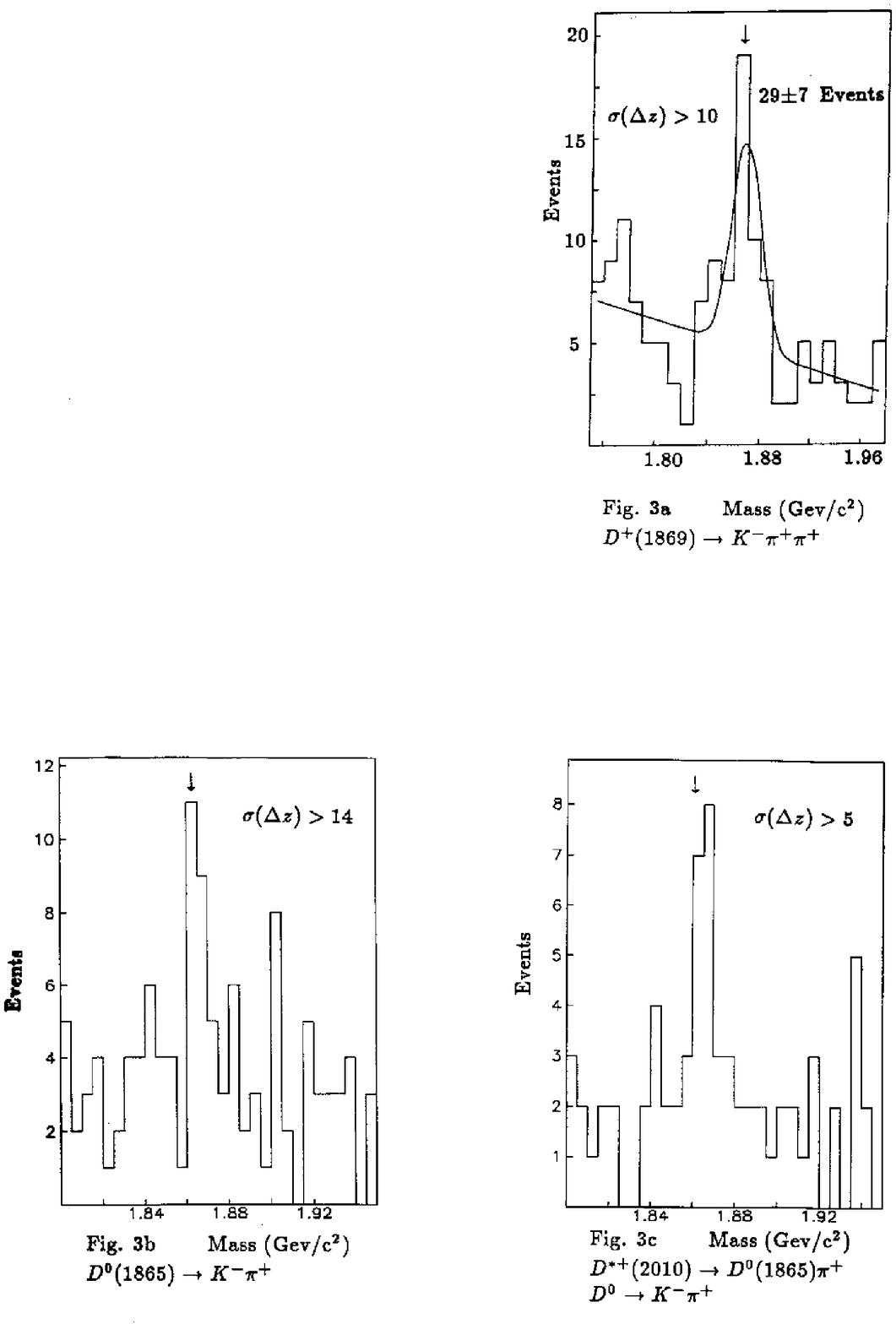}}}
\vspace*{-219mm}
\leftline{\bf REFERENCES}
%
\def\issue(#1,#2,#3){\space$\underline{#1}$\space(#2)\space#3}
\def\PR(#1,#2,#3){ Phys. Rev. \issue(#1,#2,#3)}
\def\NIM(#1,#2,#3){ Nucl. Inst. Meth.\issue(#1,#2,#3)}
\def\IEEE(#1,#2,#3){IEEE Trans. Nucl. Sci.\issue(#1,#2,#3)}
\newcounter{bean}
\begin{list}
{[\arabic{bean}]}{\usecounter{bean} \setlength{\leftmargin}{6.0mm}
\setlength{\rightmargin}{68.0mm}
\setlength{\itemsep}{0pt}
\setlength{\topsep}{1pt}
}
\item{J.R. Raab et al., Measurement of the $D^+$, $D^0$, and $D_{\sss}^{\lp}$
Lifetimes, \, \PR(D37,1988,2391).}
\item{T. Nash et al., The ACP Multiprocessor System at Fermilab,
   Proc. of the XXIII Int'l. Conf. on High
   Energy Physics, Berkeley (1986) 1459.}
\item{P. Karchin et al., \IEEE(32,1985,612).}
\item{G. Gidal et al., Major Detectors in Elementary Particle Physics,
     LBL-91 (May 1985);
     D.~Bartlett et al., \NIM(A260,1987,55);
     V.K.~Bharadwaj et al.,\NIM(155,1978,411);
     V.K.~Bharadwaj et al.,\NIM(228,1985,283);
     D.J.~Summers,\NIM(228,1985,290);
     J.A.~Appel et al., \NIM(A243,1986,361).}
\item{C. Gay and S. Bracker, \IEEE(34,1987,870);
     R. Vignani et al., \IEEE(34,1987,756);
     S. Hansen et al., \IEEE(34,1987,1003);
     Mark Bernett et al., \IEEE(34,1987,1047).}
\item{R. Hance et al., \IEEE(34,1987,878).}
%
%
\end{list}
\end{document}